\documentclass[runningheads]{llncs}
\usepackage{graphicx}
\usepackage{amssymb}
\usepackage{amsmath}
\usepackage{bm} 

\usepackage[T1]{fontenc}
\usepackage{url}
\usepackage{fixltx2e}
\usepackage{graphicx}
\usepackage{multirow}
\usepackage{multicol}
\usepackage{booktabs}
\usepackage{orcidlink}
\usepackage[ruled,vlined]{algorithm2e}   % reader-friendly pseudocode
\usepackage{subcaption}
\usepackage{hyperref}
\hypersetup{
  colorlinks=true,
  linkcolor=blue,     % For references
  citecolor=green     % For citations
}

\begin{document}
\mainmatter              % start of a contribution
\title{PoPStat–COVID19: Leveraging Population Pyramids to Quantify Demographic Vulnerability to COVID-19}
\titlerunning{PoPStat-COVID19}  % abbreviated title (for running head)
% %                                     also used for the TOC unless
% %                                     \toctitle is used
% %
% \author{Ivar Ekeland\inst{1} \and Roger Temam\inst{2}
% Jeffrey Dean \and David Grove \and Craig Chambers \and Kim~B.~Bruce \and
% Elsa Bertino}

\author{Buddhi Wijenayake\inst{1*}\orcidlink{0009-0001-2624-0251} \and Athulya Ratnayake\inst{1}\orcidlink{0009-0008-9582-4606}\and
Lelumi Edirisinghe\inst{2}\orcidlink{0009-0009-5113-2525}\and Uditha Wijeratne\inst{3}\orcidlink{0009-0006-8108-9490} \and Tharaka Fonseka \inst{4}\orcidlink{0009-0005-6403-0166}\and
Roshan Godaliyadda\inst{1}\orcidlink{0000-0002-3495-481X}\and
Samath Dharmaratne\inst{5}\orcidlink{0000-0003-4144-2107}\and Parakrama Ekanayake
\inst{1}\orcidlink{0000-0002-5639-8105} \and Vijitha Herath\inst{1}\orcidlink{0000-0002-2094-0716}\and
Inosha Alwis\inst{5,6}\orcidlink{0000-0001-6547-7449}\and Supun Manathunga\inst{7}\orcidlink{0000-0002-2607-3430}}

% %
\authorrunning{B. Wijenayake et al.} % abbreviated author list (for running head)

\institute{Department of Electrical and Electronic Engineering, Faculty of Engineering, University of Peradeniya, Peradeniya, Sri Lanka\and Department of Mathematics, Faculty of Science, University of Ruhuna, Sri Lanka \and Faculty of Medicine, University of Peradeniya, Sri Lanka\and Multidisciplinary AI Research Center, University of Peradeniya, Peradeniya, Sri Lanka \and Department of Community Medicine, Faculty of Medicine, University of Peradeniya, Peradeniya, Sri Lanka \and Australian Centre for Health Services Innovation (AusHSI) and Centre for Healthcare Transformation, School of Public Health and Social Work, Queensland University of Technology (QUT), Brisbane, Queensland, Australia \and Division of Experimental Medicine, Department of Medicine, Faculty of Medicine, McGill University, Canada.
\\
\inst{*}Corresponding Author, email:\email{e19445@eng.pdn.ac.lk}}

% %
% %%%% list of authors for the TOC (use if author list has to be modified)
% \tocauthor{Ivar Ekeland, Roger Temam, Jeffrey Dean, David Grove,
% Craig Chambers, Kim B. Bruce, and Elisa Bertino}
% %
% \institute{Princeton University, Princeton NJ 08544, USA,\\
% \email{I.Ekeland@princeton.edu},\\ WWW home page:
% \texttt{http://users/\homedir iekeland/web/welcome.html}
% \and
% Universit\'{e} de Paris-Sud,
% Laboratoire d'Analyse Num\'{e}rique, B\^{a}timent 425,\\
% F-91405 Orsay Cedex, France}

\maketitle              % typeset the title of the contribution

\begin{abstract}
Understanding how population age structure shapes COVID-19 burden is crucial for pandemic preparedness, yet common summary measures such as \emph{median age} ignore key distributional features like skewness, bimodality, and the proportional weight of high-risk cohorts. We extend the PoPStat framework, originally devised to link entire population pyramids with cause-specific mortality by applying it to COVID-19. Using 2019 United Nations World Population Prospects age–sex distributions together with cumulative cases and deaths per million recorded up to 5~May~2023 by Our World in Data, we calculate \emph{PoPDivergence} (the Kullback–Leibler divergence from an optimised reference pyramid) for 180+ countries and derive \emph{PoPStat–COVID19} as the Pearson correlation between that divergence and log-transformed incidence or mortality. Optimisation selects Malta’s old-skewed pyramid as the reference, yielding strong negative correlations for cases (\(r=-0.86\), \(p<0.001\), \(R^{2}=0.74\)) and deaths (\(r=-0.82\), \(p<0.001\), \(R^{2}=0.67\)). Sensitivity tests across twenty additional, similarly old-skewed references confirm that these associations are robust to reference choice. Benchmarking against eight standard indicators like gross domestic product per capita, Gini index, Human Development Index, life expectancy at birth, median age, population density, Socio-demographic Index, and Universal Health Coverage Index shows that \emph{PoPStat–COVID19} surpasses GDP per capita, median age, population density, and several other traditional measures, and outperforms every comparator for fatality burden. \emph{PoPStat–COVID19} therefore provides a concise, distribution aware scalar for quantifying demographic vulnerability to COVID-19.
\end{abstract}

\section{Introduction}

COVID-19, a highly contagious and plausibly severe respiratory disease caused by the SARS-CoV-2 virus~\cite{C_Huang_2020}, has yielded global mortality of approximately 7.1 million confirmed deaths~\cite{who_covid19_2025}, persisting as a significant public health threat despite the end of its formal emergency phase. Declared a Public Health Emergency of International Concern (PHEIC) by the WHO on January 30, 2020, and a pandemic on March 11, 2020, the acute emergency status ended on May 5, 2023~\cite{WHO2025news}, when the cumulative number of deaths reached approximately 6.9 million~\cite{who_covid19_2025}. Nevertheless, the virus continues to circulate, and by May 25, 2025, the confirmed global death count had risen to around 7.1 million~\cite{who_covid19_2025}. The severity of COVID-19, as reflected in both cases and fatalities, has varied significantly across countries; however, its overall impact underscores its persistent global health risk.

Older age is the strongest predictor of COVID-19 severity~\cite{C_Zhou_2020,S_undefined_2020,R_Wu_2020,R_Chen_2020}, as evidenced by significantly higher case-fatality rates (CFRs) and mortality risk ratios in older cohorts. US and Italian studies documented CFRs above 19--27\% for those aged $\geq$80 years, conflicting sharply with rates below 1\% for younger adults aged 20--54~\cite{S_undefined_2020,C_Livingston_2020}.Further supporting this,~\cite{R_Wu_2020} reported a hazard ratio for mortality greater than 6 for older populations. Similarly, \cite{Lin_Fu_2020} reported a relative risk ratio greater than 10 for older populations.

Additional studies from multiple regions support the significant role of demographics in COVID‑19 outcomes. For example, \cite{Dudel_2020} found that more than half of the variation in case‐fatality rates (CFRs) between countries with low and high CFR can be explained by differences in age structure. In \cite{Goldstein_2020}, the patterns of COVID-19 deaths by age in many countries closely reflected the rise in all-cause deaths as people got older. This shows that older age groups carry a heavier burden. \cite{Sudharsanan_2020} found that after adjusting for the age distribution of cases, up to 66\% of differences in CFRs between countries could be explained. This resulted in an age-standardized median CFR of about 1.9\%. More recently (2024), \cite{Zhou_2024} created a global machine-learning model covering 156 countries. They found that in "ageing-driven" clusters, which mainly include high-income European countries, demographic factors such as the proportion of elderly people significantly influenced higher CFRs. In sub-national analyses, \cite{Perone_2021} conducted in Italian provinces showed that provinces with larger populations aged 70 and older, particularly those with many people aged 80 to 90, had significantly higher case fatality rates. This was especially true when the healthcare system was under strain and environmental pollution levels were high.

Previous work has linked simple population‐pyramid categories to COVID-19 outcomes. For example, classifying countries as “expansive” (young) or “constrictive” (old) shows that aging pyramids often have higher mortality~\cite{D_Manuel_2020}, but this binary approach cannot capture gradual shifts in age structure. Other studies include the pyramid shape in multivariate models but cannot disentangle demographic effects from healthcare or socioeconomic factors~\cite{C_Birchenall-Jiménez_2024}. A single‐country analysis in Pakistan attributes low mortality to its youthful age distribution~\cite{A_Khan_2022}, yet its conclusions may not generalize. These descriptive methods lack a unified, scalable metric for quantifying how the entire population pyramid relates to COVID-19 burden—a gap that PoPStat-COVID19 is designed to fill.  

The median age, another common demographic summary statistic used in analyses like those of \cite{A_Wang_2020}, provides an incomplete picture of population vulnerability, as it fails to capture the full structural complexity and shape of the age distribution. While offering some contextual value, its limitation lies in ignoring critical features, such as skewness, bimodality, or the proportional weight of high-risk age groups within a population pyramid, which can potentially overlook the granular demographic risk profile relevant to COVID-19 mortality.

PoPStat, introduced in \cite{popstat}, overcomes these limitations of single‐value summaries by leveraging the full age–sex distribution of a population.  It quantifies demographic vulnerability by computing the Pearson correlation between the Kullback–Leibler divergence of each country's population pyramid from an optimized reference pyramid, and the corresponding cause-specific mortality rates observed in those countries.  In global analysis \cite{popstat}, covering 371 disease categories, PoPStat explained over 80\% of cross‐national variation in mortality for neurological disorders (\(r=0.90\)) and neoplasms (\(r=-0.90\)), outperforming traditional socio‐economic and demographic indicators.

However, PoPStat has not been applied to cumulative COVID-19 outcomes.  In this study, we adapt PoPStat to the COVID-19 pandemic—denoted PoPStat–COVID19 and address three primary objectives,
\begin{enumerate}
  \item \textbf{Construction of PoPStat-COVID19}  
    We derive \(\text{PoPStat}-{\text{COVID19}}\) by selecting an optimized reference pyramid (Malta) that maximizes correlation with both COVID-19 incidence and mortality to compute the scalar association for both cases and deaths per million.  
    \vspace{0.5em}
  \item \textbf{Robustness analysis across reference choices.}We assess the stability of PoPStat–COVID19 by conducting sensitivity analyses over twenty additional high‐performing reference pyramids, showing that similarly skewed age structures yield similar correlation coefficients for cases and deaths.

  \vspace{0.5em}
  \item \textbf{Benchmarking against standard metrics.}  
    We compare PoPStat–COVID19 to gross domestic product(GDP) per capita, Gini index of income inequality, Human Development Index(HDI), life expectancy at birth, population median age, population density, Socio-demographic Index(SDI) and Universal Health Coverage Index(UHCI), showing that the demographic–structure signal rivals these conventional predictors in explaining COVID-19 burden.
\end{enumerate}

\section{Methods}
\subsection{Preliminaries}
The metrics PoPStat and PoPDivergence, proposed in \cite{popstat}, enable the conversion of multidimensional population structures into scalar variables, facilitating a quantitative analysis of the relationship between population demographics and disease-specific mortality. PoP-Divergence is a Kullback-Leibler (KL) divergence
based measure to quantify the structural deviation of a population pyramid from an optimized reference population, as shown in \eqref{div}.

\begin{equation}
  \mathrm{PoPDivergence}(i;\,\omega)
  = \sum_{a\in A} P_{i,a}\,\ln\!\left(\frac{P_{i,a}}{P_{\omega,a}}\right)
  \label{div}
\end{equation}

where $P_{i,a}$ denotes the proportion of the population in age group $a$ for country $i$, and $P_{\omega,a}$ denotes the corresponding proportion in the reference country $\omega$.  The index $ a\in A$ spans all age groups, and captures the information‐theoretic divergence of country $i$ ’s pyramid from that of reference $\omega$, with larger values indicating greater structural deviation \cite{popstat}.

\begin{figure}[htbp]
    \centering
    \hspace*{-2cm}
    \begin{tabular}{@{}cccccc@{}}
        {\footnotesize PoPDivergence} & 0 & 0.009 & 0.148 & 0.412 & 0.564 \\[1ex]
         & \includegraphics[width=0.18\textwidth]{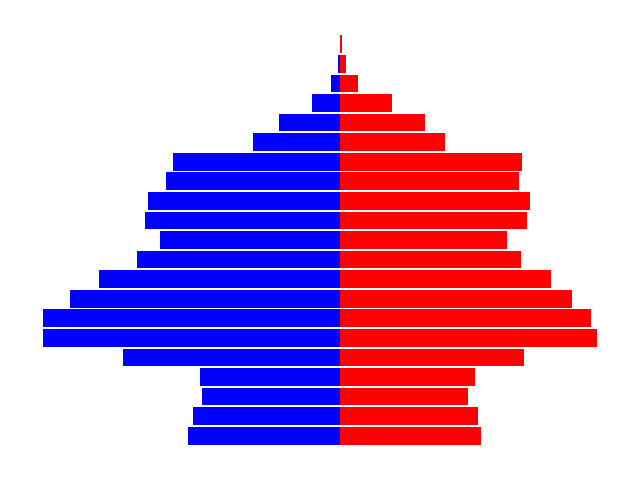} 
         & \includegraphics[width=0.18\textwidth]{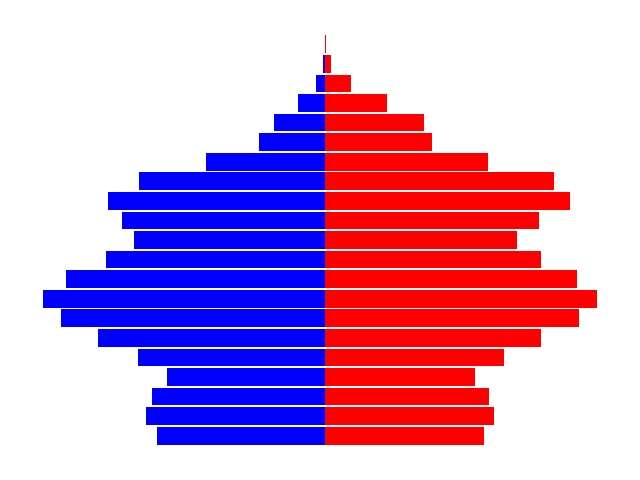} 
         & \includegraphics[width=0.18\textwidth]{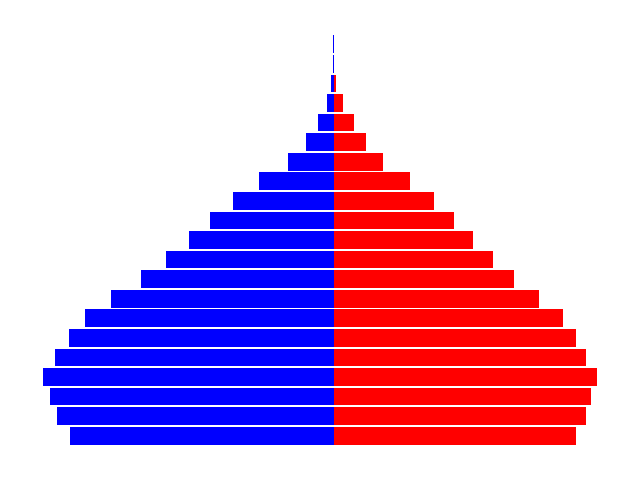} 
         & \includegraphics[width=0.18\textwidth]{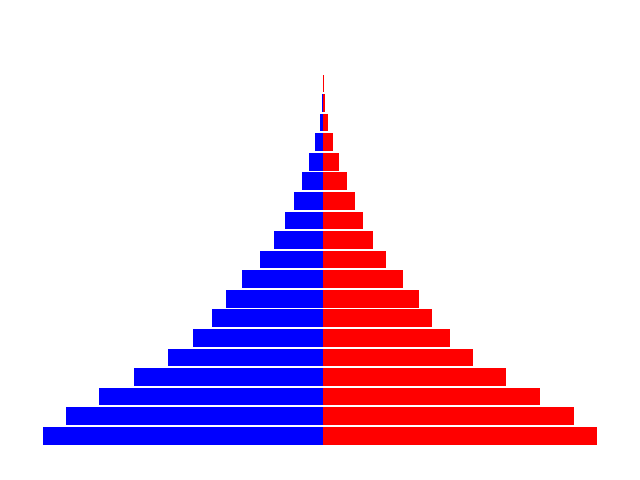} 
         & \includegraphics[width=0.18\textwidth]{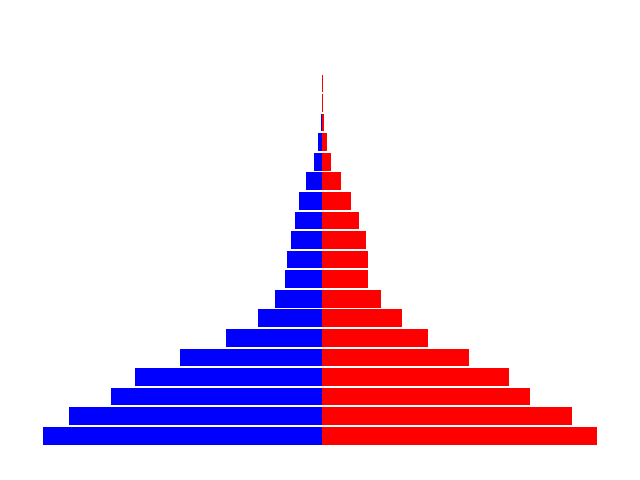} \\[1ex]
         & Malta & Poland & India & Nigeria & Central African Rep. \\
    \end{tabular}
    \caption{Deviation of PoPDivergence. Malta as the reference, indicating movement toward progressive structures.}
    \label{fig:pop_div}
\end{figure}    

Moreover, the interpretation of \textit{PoPDivergence} depends critically on the choice of reference population. As illustrated in Figure \ref{fig:pop_div}, selecting Malta—a country with a comparatively regressive age structure—as the reference causes higher \textit{PoPDivergence} values to correspond to progressively younger (more expansive) pyramids relative to Malta. The optimal reference country is identified via reference‐country tuning \cite{popstat}, which chooses the reference that maximizes the Pearson correlation between \textit{PoPDivergence} and the natural logarithm of the cause‐specific mortality rate. Formally, this selection solves

\begin{equation}
  \omega^*
  \;=\;
  \underset{\omega\in\Omega}{\arg\max}
  \;\bigl|\mathrm{Cor}\bigl[\{\ln S_i\}_{i\in\Omega},\;\{\mathrm{PoPDivergence}(i;\omega)\}_{i\in\Omega}\bigr]\bigr|
  \label{objective}
\end{equation}

Here \(\Omega\) denotes the set of all countries and \(S_i\) the crude death rate in country \(i\).  The notation \(\{\mathrm{PoPDivergence}(i;\omega)\}_{i\in\Omega}\) indicates the vector of divergence values—one per country—computed against reference \(\omega\), while \(\{\ln S_i\}_{i\in\Omega}\) is the vector of log‐transformed crude death rates. Then \(\omega\) that maximizes the Pearson correlation between those two vectors, is selected. This ensures the selected reference yields the strongest linear association between demographic divergence and mortality patterns.

PoPStat serves as a scalar summary of the linear association between demographic divergence and mortality.  Once the optimal reference population \(\omega^*\) has been selected via Equation \eqref{objective}, we calculate PoPStat as in \eqref{popstat-eq}.

\begin{equation}
    PoPStat(ref:\omega^*) = \frac{\sum_{i=1}^{n} (x_i - \bar{x})(y_i - \bar{y})}{\sqrt{\sum_{i=1}^{n} (x_i - \bar{x})^2 \sum_{i=1}^{n} (y_i - \bar{y})^2}}
    \label{popstat-eq}
\end{equation}

where \( x_i \) and \( y_i \) represent the PoPDivergence and the natural logarithm of the mortality rates for the \( i \)-th country, respectively, and \( \bar{x} \) and \( \bar{y} \) are the mean values of the vectors \( X \) (PoPDivergence values) and \( Y \) (log-transformed mortality rates). The resulting PoPStat value reflects how strongly the population structure influences mortality patterns for specific diseases. Higher absolute values of PoPStat indicate a stronger association between the demographic structure and disease-specific mortality.

\begin{algorithm}[t]
\footnotesize
\linespread{1.4}\selectfont  % adjust line spacing
\caption{Constructing PoPStat–COVID19}
\label{algo}
\KwIn{Population proportions $P_{i,a}$ for age‐group $a\in A$ and country $i\in\Omega$, cumulative cases $C_i$ and deaths $D_i$ per million}
\KwOut{PoPStat$-{\mathrm{COVID19_{cases}}}$, PoPStat$-{\mathrm{COVID19_{deaths}}}$}

\BlankLine
\textbf{Step 1: Compute log‐outcomes}\;
\ForEach{$i\in\Omega$}{
  $\mathrm{logCases}_i  \leftarrow \ln(C_i)$\;
  $\mathrm{logDeaths}_i \leftarrow \ln(D_i)$\;
}

\BlankLine
\textbf{Step 2: Reference–country tuning}\;
\ForEach{$\omega\in\Omega$}{
  \ForEach{$i\in\Omega$}{
    $\mathrm{PoPDivergence}(i;\,\omega)
      \leftarrow \sum_{a\in A} P_{i,a}\,\ln\!\left(\frac{P_{i,a}}{P_{\omega,a}}\right)$\;
  }
  $\rho_{\mathrm{cases}}[\omega]
     \leftarrow \bigl|\mathrm{corr}(\{\mathrm{PoPDivergence}(i;\,\omega)\}_{i\in\Omega},\{\mathrm{logCases}_i\}_{i\in\Omega})\bigr|$\;
  $\rho_{\mathrm{deaths}}[\omega]
     \leftarrow \bigl|\mathrm{corr}(\{\mathrm{PoPDivergence}(i;\,\omega)\}_{i\in\Omega},\{\mathrm{logDeaths}_i\}_{i\in\Omega})\bigr|$\;
}
$\omega_{cases}^* \leftarrow \arg\max_{\omega\in\Omega}(\rho_{\mathrm{cases}})$\;
$\omega_{deaths}^* \leftarrow \arg\max_{\omega\in\Omega}(\rho_{\mathrm{deaths}})$\;

\BlankLine
\textbf{Step 3: Compute final PoPStat–COVID19}\;
{\scriptsize
$\mathrm{PoPStat\text{-}COVID19}_{\mathrm{cases}}
  \leftarrow \mathrm{corr}(\{\mathrm{PoPDivergence}(i;\omega^*_{\mathrm{cases}})\}_{i\in\Omega},\{\mathrm{logCases}_i\}_{i\in\Omega})$\;
$\mathrm{PoPStat\text{-}COVID19}_{\mathrm{deaths}}
  \leftarrow \mathrm{corr}(\{\mathrm{PoPDivergence}(i;\omega^*_{\mathrm{deaths}})\}_{i\in\Omega},\{\mathrm{logDeaths}_i\}_{i\in\Omega})$\;
}

\end{algorithm}

\subsection{Data Collection and Preprocessing}
\label{data-collect}
We obtained 183 country‐level population estimates for the year 2019 from the United Nations World Population Prospects 2024, choosing 2019 because it reflects the demographic structure immediately prior to the global emergence of SARS-CoV-2 in late 2019. From these data, we constructed country‐specific population pyramids in five-year age groups and normalized them to gender‐specific proportions. COVID-19 cases and deaths for each country were retrieved from Our World in Data \cite{owid-coronavirus}, aggregated up to 5 May 2023, the date on which the World Health Organization declared COVID-19 no longer a pandemic~\cite{WHO2025news}.  

\subsection{Constructing PoPStat-COVID19}

To construct PoPStat–COVID19, we merged the 2019 population pyramids with the cumulative COVID-19 surveillance data outlined in Section \ref{data-collect} and applied the reference-country tuning procedure in Equation \ref{objective}. Every country $\omega\in\Omega$ was treated as a candidate reference pyramid for the two log-transformed severity measures, cumulative cases per million and cumulative deaths per million. For each candidate reference pyramid we computed PoPDivergence for all countries and then calculated the Pearson correlation between PoPDivergences and the severity measure. The pyramid that maximised the absolute correlation was selected as the optimal reference country $\omega^*$. 

Since \textbf{Malta} emerged as optimal reference country $\omega^*$ for both scenarios, using Malta as the reference we derived \textbf{PoPStat–COVID19} by correlating those divergences with each outcome according to Equation \eqref{popstat-eq}. The complete algorithm is outlined in the Algorithm \ref{algo}.

\subsection{Impact of Reference Pyramid on PoPStat–COVID19}
\label{imapact-ref}

To assess robustness to the choice of reference, we applied the tuning procedure in \ref{objective} to identify the ten candidate pyramids yielding the most extreme negative correlations with log‐COVID-19 deaths per million and the ten yielding the most extreme positive correlations. For each of these twenty reference populations, PoPStat–COVID19 was then recalculated for both cumulative cases and cumulative deaths per million.

\raggedbottom
\begin{figure}[t]
    \centering
    \makebox[\textwidth][c]{%
        \begin{subfigure}[t]{0.5\textwidth}
            \centering
            \includegraphics[width=\textwidth]{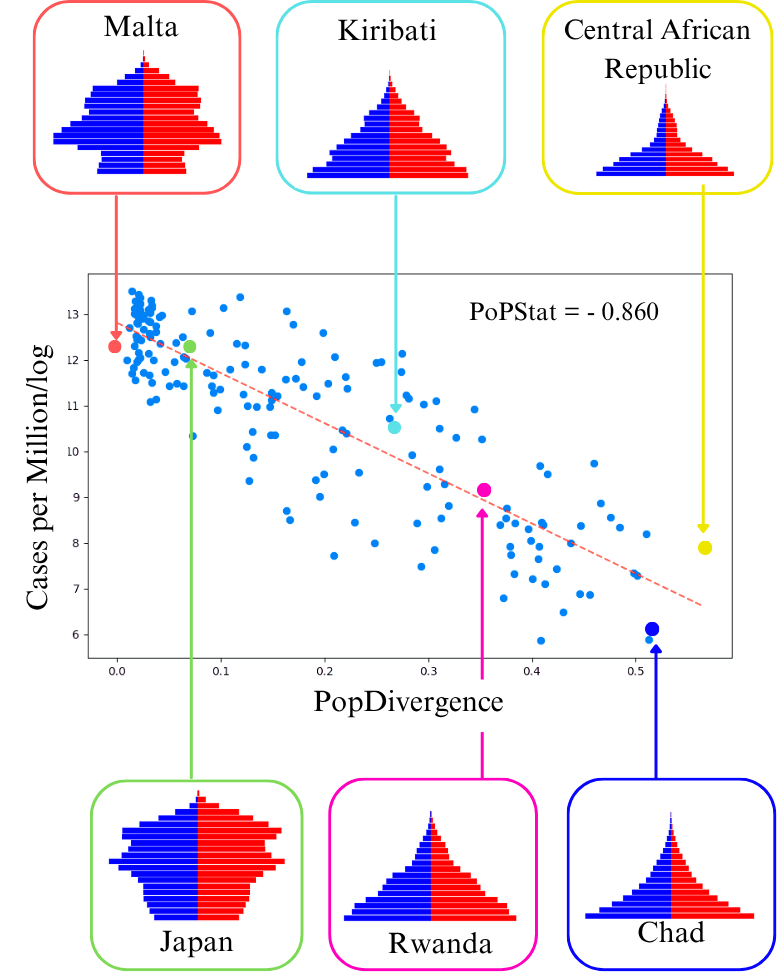}
            \caption{COVID-19 Cases}
            \label{fig:cases_pop}
        \end{subfigure}
        \hspace{2em}
        \begin{subfigure}[t]{0.5\textwidth}
            \centering
            \includegraphics[width=\textwidth]{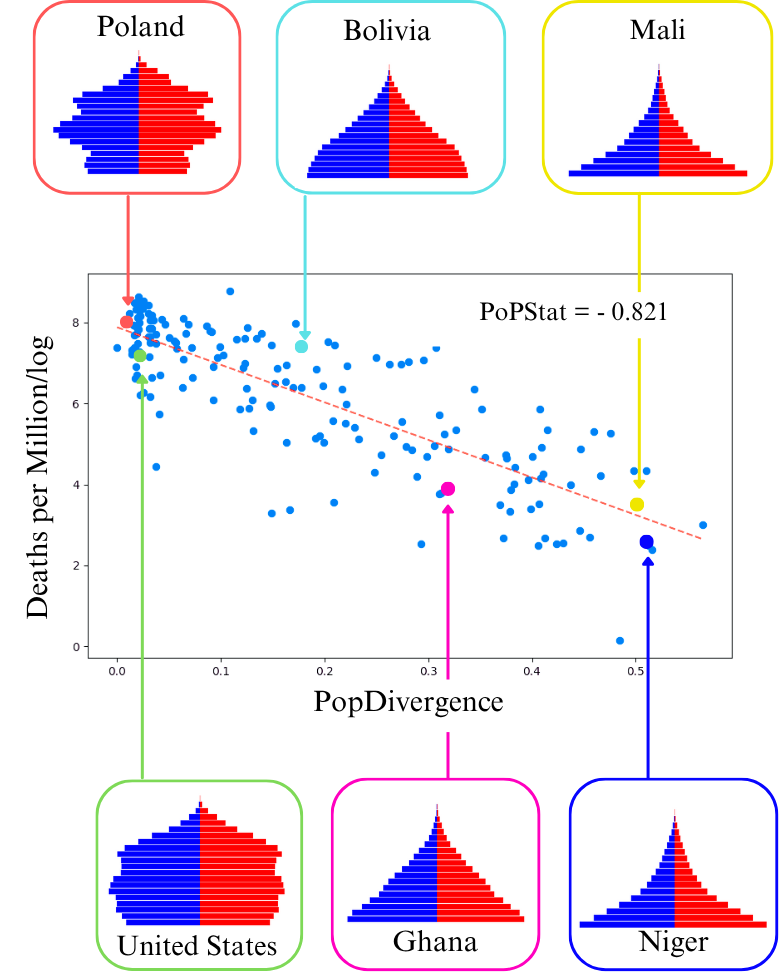}
            \caption{COVID-19 Deaths}
            \label{fig:deaths_pop}
        \end{subfigure}
    }

    \caption{PoPStat-COVID19 for COVID-19 Cases and Deaths with Malta as the reference population pyramid}
    \label{fig:popstat_cases_deaths}
\end{figure}
 
\subsection{Comparison of PoPStat-COVID19 with other indicators}
To benchmark PoPStat–COVID19 against established development and demographic metrics, we compiled country‐level indicators for 2019, namely GDP per capita~\cite{worldbank_gdp_2024}, Gini index~\cite{worldbank_gini_2024}, Human Development Index (HDI)~\cite{undp_hdr_datacenter_2024}, life expectancy at birth~\cite{who_life_expectancy_2024}, median age~\cite{un_wpp_2024}, population density~\cite{worldbank_pop_density_2024}, Socio‐demographic Index (SDI)~\cite{ihme_gbd_2024}, and the Universal Health Coverage Index (UHCI)~\cite{noauthor_universal_nodate}. For all indicators except the Gini index, data were available for the full set of 183 countries $\Omega$ used in PoPStat–COVID19; Gini index values could only be retrieved for 95 countries. We then computed Pearson correlation coefficients between COVID-19 cases and deaths and each socio‐economic indicator, and fitted simple linear regressions to compare the proportion of variance explained ($R^2$) by PoPStat–COVID19 relative to these traditional covariates.

\section{Results}

\newcommand{\popstat}[1]{PoPStat-COVID19_{#1}}
\subsection{Association of PoPStat–COVID19 with COVID-19 Burden}

Figure \ref{fig:popstat_cases_deaths} illustrates how the COVID‐19 burden varies with demographic divergence from Malta’s population pyramid.  In the left panel, the $\popstat{cases}$ (the Pearson correlation between PoPDivergence and cases per million) is \(-0.860\) (\(p<0.001\)), indicating that countries whose age structures differ more from Malta tend to have substantially fewer reported cases.  In the right panel, the $\popstat{deaths}$ (the Pearson correlation between PoPDivergence and deaths per million) is \(r=-0.821\) (\(p<0.001\)), confirming a similarly strong negative association for mortality.  

\subsection{Robustness to Reference–Population Choice}
\label{sec:consistency_refs}
\begin{table*}[t]
\centering
\scriptsize
\setlength{\tabcolsep}{4pt}
\renewcommand{\arraystretch}{1.25} % add vertical breathing room
\caption{PoPStat–COVID19 robustness across reference pyramids}
\label{tab:consistency}
\begin{tabular*}{\textwidth}{@{\extracolsep{\fill}} l ccc ccc}
\toprule
\multirow{2}{*}{Reference} & \multicolumn{3}{c}{Cases} & \multicolumn{3}{c}{Deaths}\\
\cmidrule(lr){2-4}\cmidrule(lr){5-7}
 & PoPStat & $p$ & 95\% CI & PoPStat & $p$ & 95\% CI\\
\midrule
\multicolumn{7}{l}{\textbf{Regressive (old-skewed) references}}\\
Malta              & --0.860 & $<0.001$ & (--0.89, --0.82) & --0.821 & $<0.001$ & (--0.86, --0.77)\\
Poland             & --0.852 & $<0.001$ & (--0.89, --0.81) & --0.819 & $<0.001$ & (--0.86, --0.76)\\
Ukraine            & --0.853 & $<0.001$ & (--0.89, --0.81) & --0.818 & $<0.001$ & (--0.86, --0.76)\\
Hungary            & --0.849 & $<0.001$ & (--0.89, --0.80) & --0.818 & $<0.001$ & (--0.86, --0.76)\\
Slovakia           & --0.849 & $<0.001$ & (--0.89, --0.80) & --0.818 & $<0.001$ & (--0.86, --0.76)\\
Switzerland        & --0.850 & $<0.001$ & (--0.89, --0.80) & --0.817 & $<0.001$ & (--0.86, --0.76)\\
Austria            & --0.850 & $<0.001$ & (--0.89, --0.80) & --0.817 & $<0.001$ & (--0.86, --0.76)\\
Estonia            & --0.847 & $<0.001$ & (--0.88, --0.80) & --0.816 & $<0.001$ & (--0.86, --0.76)\\
Serbia             & --0.846 & $<0.001$ & (--0.88, --0.80) & --0.816 & $<0.001$ & (--0.86, --0.76)\\
\addlinespace[2pt]
\multicolumn{7}{l}{\textbf{Progressive (young-skewed) references}}\\
Central African Republic & 0.775 & $<0.001$ & (0.71, 0.83) & 0.701 & $<0.001$ & (0.62, 0.77)\\
Uganda             & 0.764 & $<0.001$ & (0.69, 0.82) & 0.697 & $<0.001$ & (0.61, 0.77)\\
Chad               & 0.764 & $<0.001$ & (0.69, 0.82) & 0.696 & $<0.001$ & (0.61, 0.77)\\
Niger              & 0.767 & $<0.001$ & (0.70, 0.82) & 0.694 & $<0.001$ & (0.61, 0.76)\\
Mali               & 0.764 & $<0.001$ & (0.69, 0.82) & 0.694 & $<0.001$ & (0.61, 0.76)\\
Somalia            & 0.764 & $<0.001$ & (0.69, 0.82) & 0.692 & $<0.001$ & (0.61, 0.76)\\
Burundi            & 0.757 & $<0.001$ & (0.69, 0.81) & 0.690 & $<0.001$ & (0.60, 0.76)\\
Afghanistan        & 0.756 & $<0.001$ & (0.69, 0.81) & 0.687 & $<0.001$ & (0.60, 0.76)\\
Mozambique         & 0.757 & $<0.001$ & (0.69, 0.81) & 0.686 & $<0.001$ & (0.60, 0.76)\\
Malawi             & 0.751 & $<0.001$ & (0.68, 0.81) & 0.684 & $<0.001$ & (0.60, 0.76)\\
\bottomrule
\end{tabular*}
\end{table*}

\begin{table*}[t]
\centering
\scriptsize
\setlength{\tabcolsep}{3pt}
\caption{Comparision of Correlation and Variance Explained by Indicators for COVID-19 Cases and Deaths vs PoPStat-COVID19}
\label{tab:popstat_covid19_full}
\begin{tabular*}{\textwidth}{@{\extracolsep{\fill}} l cccc cccc}
\toprule
Indicator
  & \multicolumn{4}{c}{Cases}
  & \multicolumn{4}{c}{Deaths} \\
\cmidrule(lr){2-5}\cmidrule(lr){6-9}
  & \(r\)   & \(p\)    & 95\% CI      & \(R^2\)
  & \(r\)   & \(p\)    & 95\% CI      & \(R^2\) \\
\midrule
GDP per capita\cite{worldbank_gdp_2024}    
  &  0.60    & $<$0.001 & ( 0.49,  0.69) & 0.36 
  &  0.38    & $<$0.001 & ( 0.24,  0.51) & 0.15 \\[0.3em]

Gini index\cite{worldbank_gini_2024}        
  & –0.33    & $<$0.001 & (–0.52, –0.11) & 0.11 
  & –0.15    & 0.21   & (–0.37,  0.08) & 0.02 \\[0.3em]

HDI\cite{undp_hdr_datacenter_2024}               
  &  0.90    & $<$0.001 & ( 0.86,  0.92) & 0.80 
  &  0.78    & $<$0.001 & ( 0.71,  0.83) & 0.61 \\[0.3em]

Life expectancy\cite{who_life_expectancy_2024}
  &  0.83    & $<$0.001 & ( 0.78,  0.87) & 0.69 
  &  0.72    & $<$0.001 & ( 0.64,  0.78) & 0.51 \\[0.3em]

Median age\cite{un_wpp_2024}        
  &  0.83    & $<$0.001 & ( 0.78,  0.87) & 0.69 
  &  0.77    & $<$0.001 & ( 0.70,  0.82) & 0.59 \\[0.3em]

Population density\cite{worldbank_pop_density_2024}
  &  0.14    & 0.07   & (–0.01,  0.28) & 0.02 
  &  0.00    & 0.96   & (–0.14,  0.15) & 0.00 \\[0.3em]

SDI\cite{ihme_gbd_2024}               
  &  0.86    & $<$0.001 & ( 0.81,  0.89) & 0.74 
  &  0.76    & $<$0.001 & ( 0.69,  0.82) & 0.57 \\[0.3em]

UHCI\cite{noauthor_universal_nodate}
  &  0.85    & $<$0.001 & ( 0.80,  0.89) & 0.72 
  &  0.78    & $<$0.001 & ( 0.72,  0.84) & 0.62 \\[0.3em] 

PoPStat-COVID19
  & \(-0.86\) & $<$0.001 & (–0.89, –0.82) & 0.74 
  & \(-0.82\) & $<$0.001 & (–0.86, –0.77) & 0.67 \\[0.3em]

\bottomrule
\end{tabular*}
\end{table*}

To verify that PoPStat–COVID19 is not an artefact of using Malta alone, we recalculated the statistic with twenty alternative references as discussed in the section \ref{imapact-ref}. The ten that produced the strongest \textit{negative} correlations with log deaths were old-skewed, regressive pyramids, and the ten that produced the strongest \textit{positive} correlations were young-skewed, progressive pyramids. Table \ref{tab:consistency} summarises the resulting coefficients for both outcomes.
\raggedbottom
\begin{figure}[t]
    \centering
    % First row: Three images
% Second row: Three more plots
    \begin{subfigure}[t]{0.325\textwidth}
        \centering
        \includegraphics[width=\textwidth]{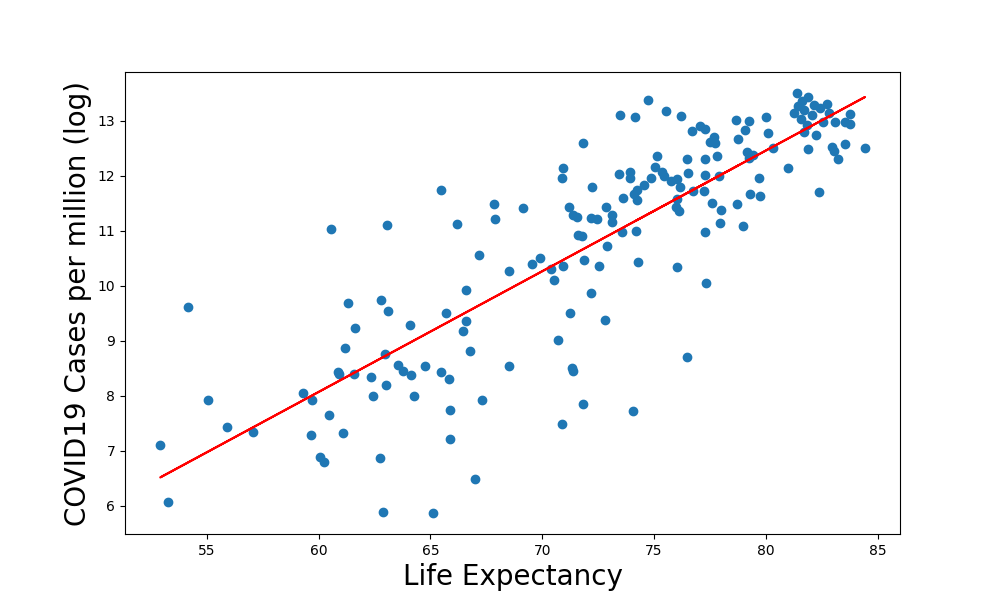}
        \caption{Life Expectancy}
        \label{fig6:plot_below}
    \end{subfigure}%
    \hfill
    \begin{subfigure}[t]{0.325\textwidth}
        \centering
        \includegraphics[width=\textwidth]{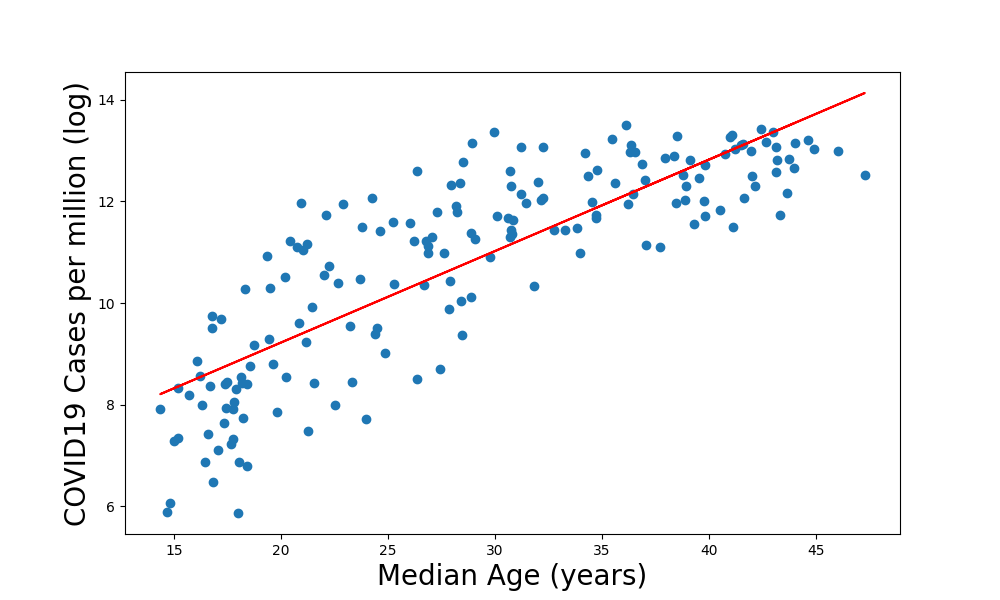}
        \caption{Median Age}
        \label{fig6:plot_below2}
    \end{subfigure}%
    \hfill
    \begin{subfigure}[t]{0.325\textwidth}
        \centering
        \includegraphics[width=\textwidth]{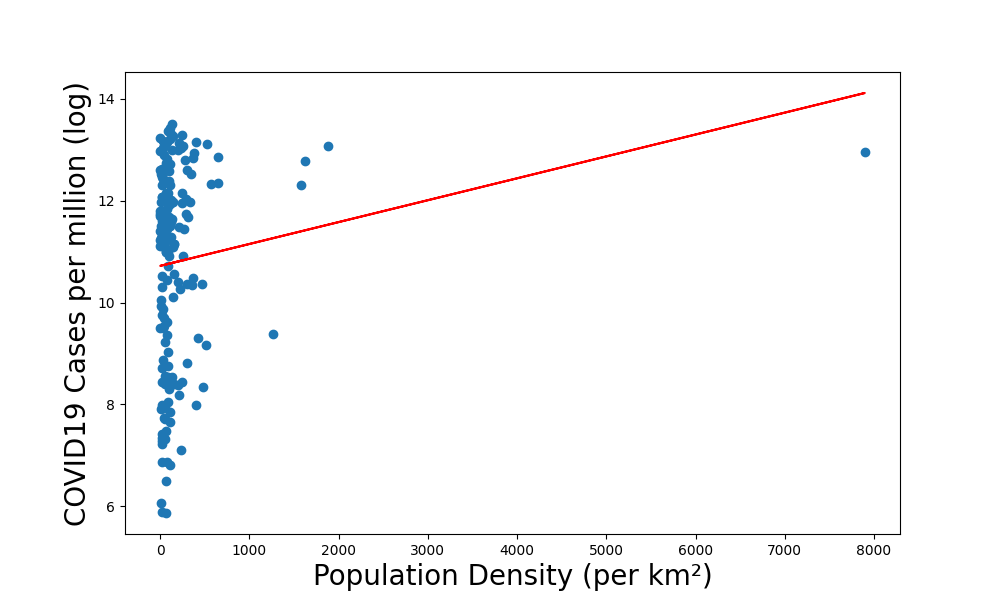}
        \caption{Population Density}
        \label{fig6:plot_below3}
    \end{subfigure}

    \vspace{1em}

    \begin{subfigure}[t]{0.325\textwidth}
        \centering
        \includegraphics[width=\textwidth]{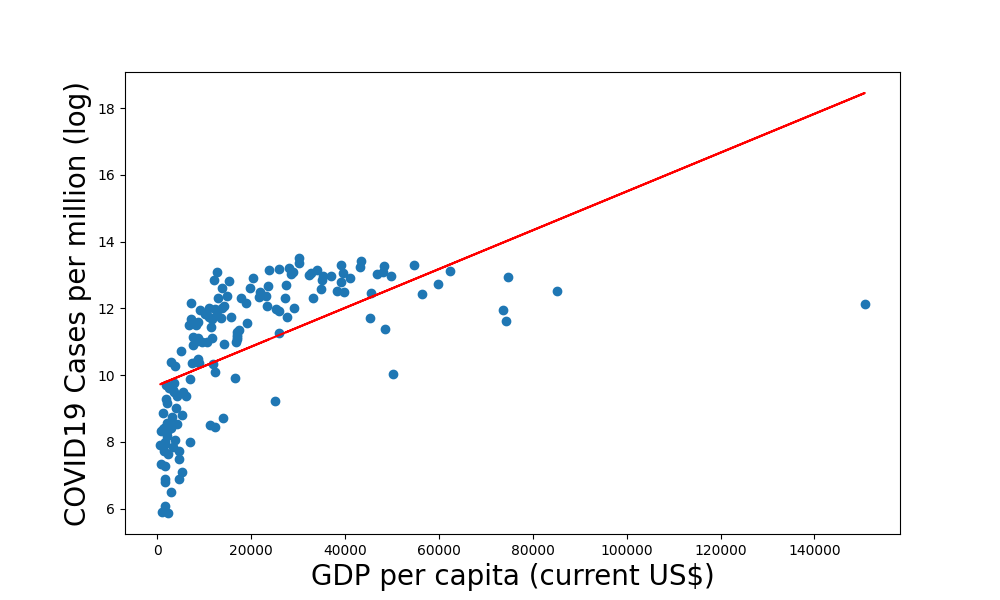}
        \caption{GDP-per-Capita}
        \label{fig6:first_image}
    \end{subfigure}%
    \hfill
    \begin{subfigure}[t]{0.325\textwidth}
        \centering
        \includegraphics[width=\textwidth]{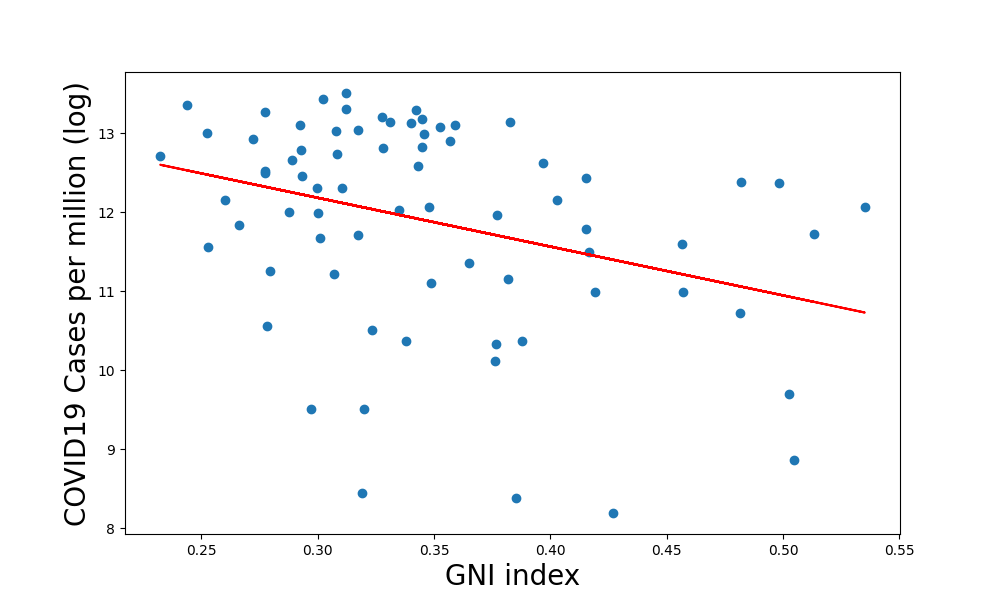}
        \caption{Gini Index}
        \label{fig6:second_image}
    \end{subfigure}%
    \hfill
    \begin{subfigure}[t]{0.325\textwidth}
        \centering
        \includegraphics[width=\textwidth]{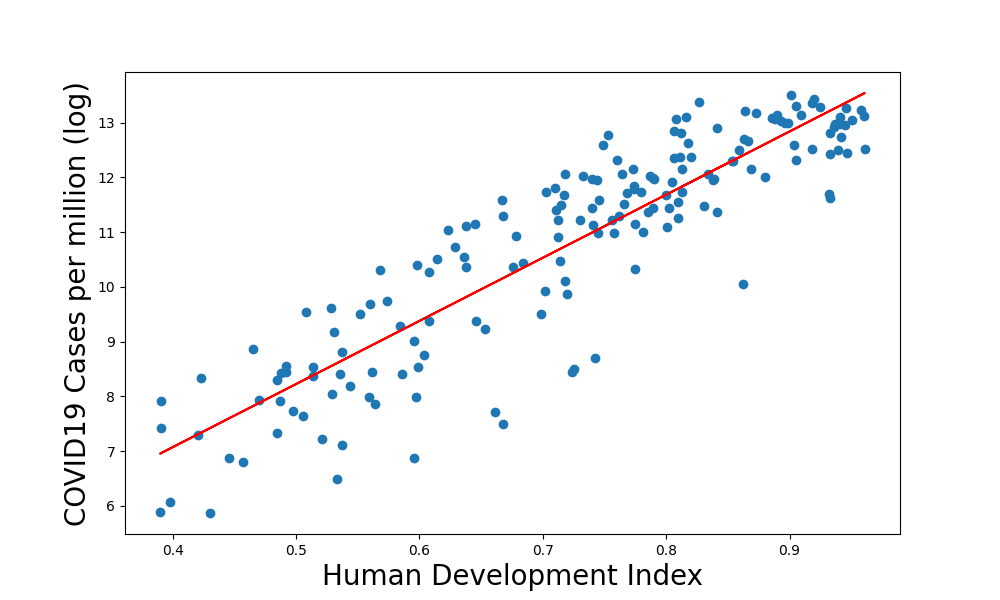}
        \caption{HDI}
        \label{fig6:third_image}
    \end{subfigure}

    \vspace{1em}

    % Third row: Two centered plots
    \makebox[\textwidth][c]{%
        \begin{subfigure}[t]{0.325\textwidth}
            \centering
            \includegraphics[width=\textwidth]{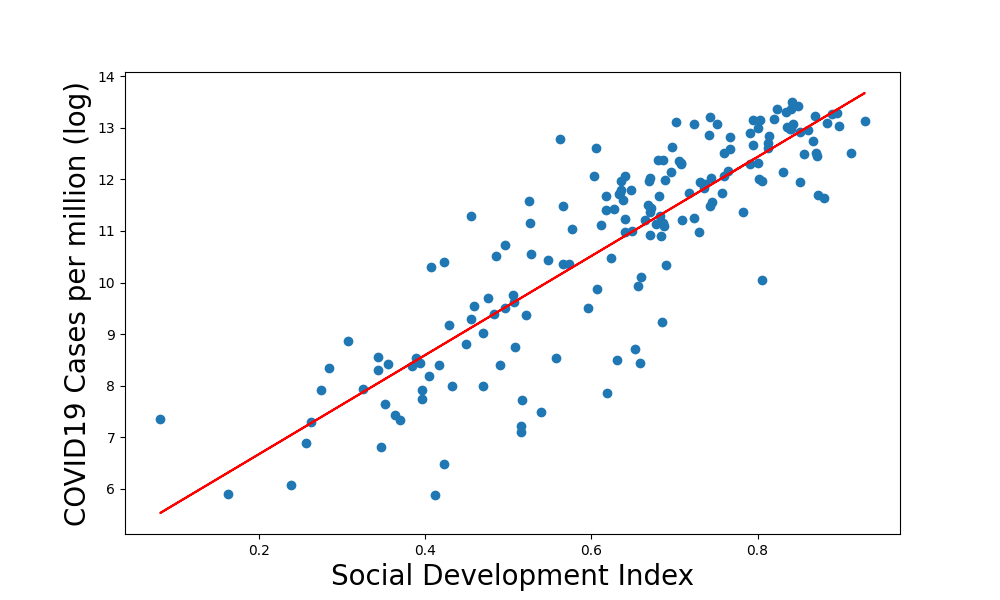}
            \caption{SDI}
            \label{fig6:plot_below4}
        \end{subfigure}
        \hspace{2em}
        \begin{subfigure}[t]{0.325\textwidth}
            \centering
            \includegraphics[width=\textwidth]{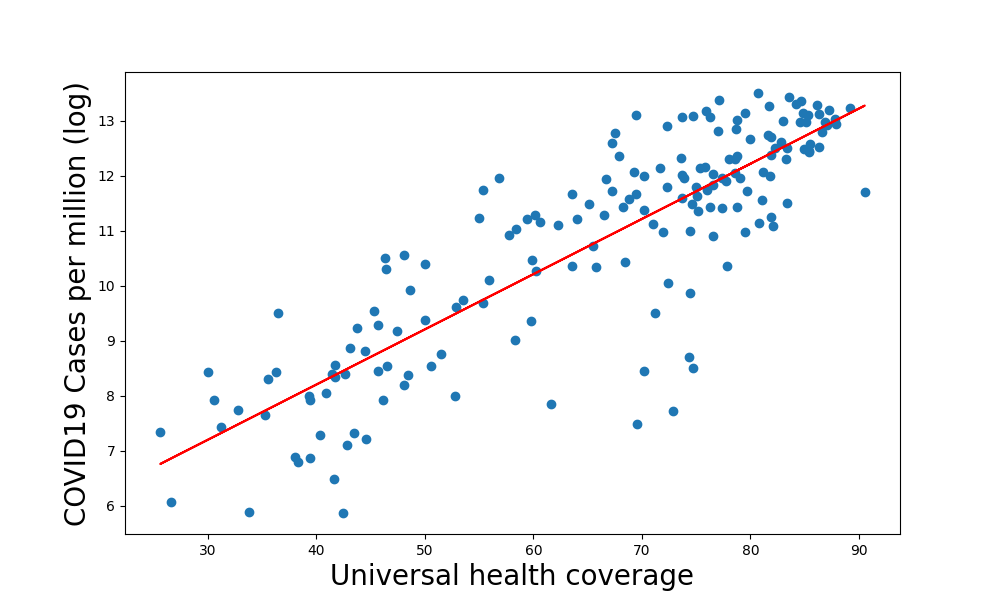}
            \caption{UHCI}
            \label{fig6:plot_below5}
        \end{subfigure}
    }

    \caption{Bivariate relationships between log–COVID-19 cases per million and conventional indicators}
    \label{fig:cases}
\end{figure}

\subsection{Benchmarking PoPStat–COVID19 against Conventional Indicators}

\begin{figure}[t]
    \centering
    % First row: Three images
% Second row: Three more plots
    \begin{subfigure}[t]{0.325\textwidth}
        \centering
        \includegraphics[width=\textwidth]{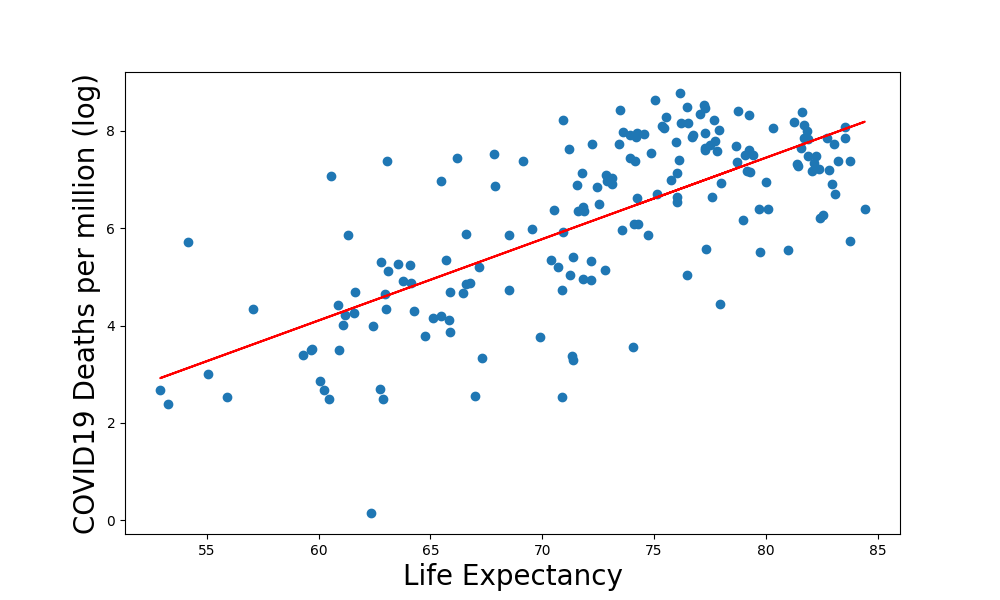}
        \caption{Life Expectancy}
        \label{fig6:plot_below}
    \end{subfigure}%
    \hfill
    \begin{subfigure}[t]{0.325\textwidth}
        \centering
        \includegraphics[width=\textwidth]{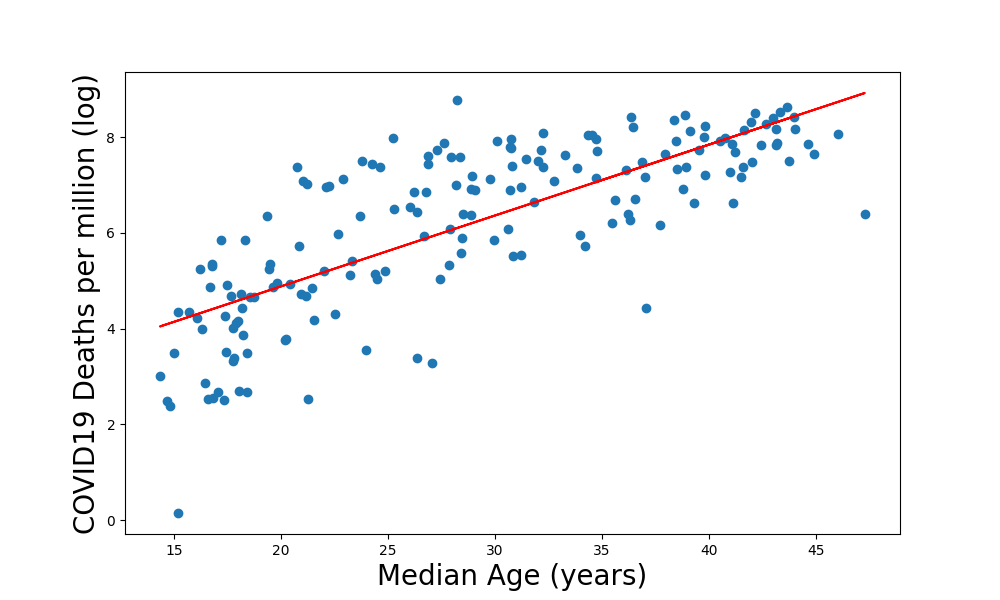}
        \caption{Median Age}
        \label{fig6:plot_below2}
    \end{subfigure}%
    \hfill
    \begin{subfigure}[t]{0.325\textwidth}
        \centering
        \includegraphics[width=\textwidth]{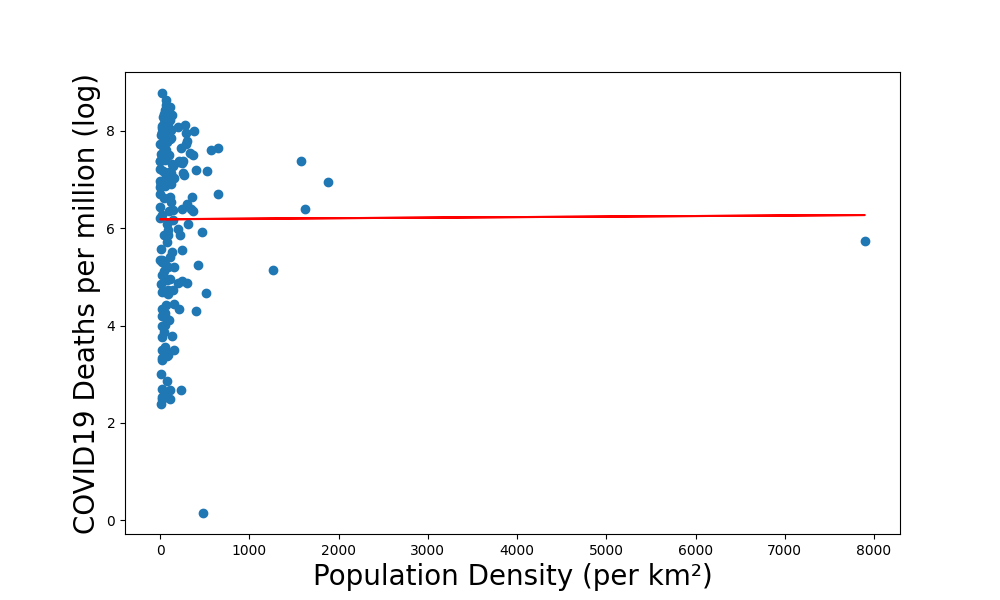}
        \caption{Population Density}
        \label{fig6:plot_below3}
    \end{subfigure}

    \vspace{1em}

    \begin{subfigure}[t]{0.325\textwidth}
        \centering
        \includegraphics[width=\textwidth]{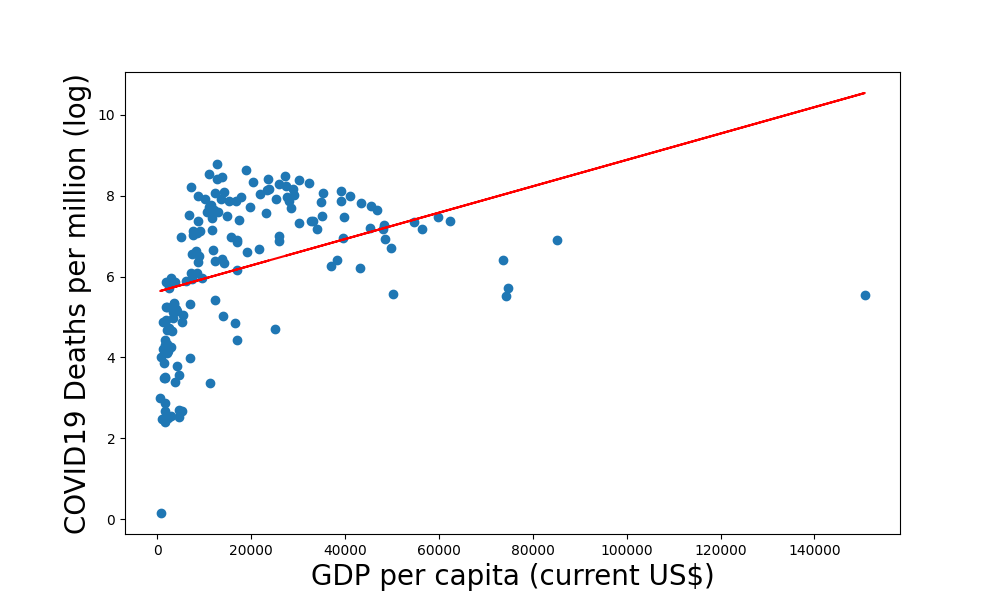}
        \caption{GDP-per-Capita}
        \label{fig6:first_image}
    \end{subfigure}%
    \hfill
    \begin{subfigure}[t]{0.325\textwidth}
        \centering
        \includegraphics[width=\textwidth]{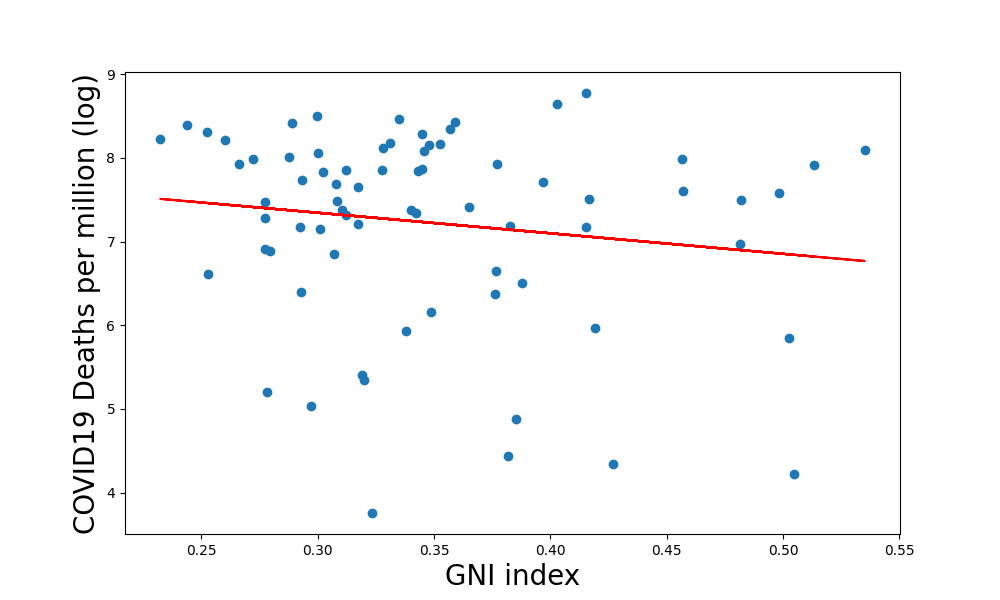}
        \caption{Gini Index}
        \label{fig6:second_image}
    \end{subfigure}%
    \hfill
    \begin{subfigure}[t]{0.325\textwidth}
        \centering
        \includegraphics[width=\textwidth]{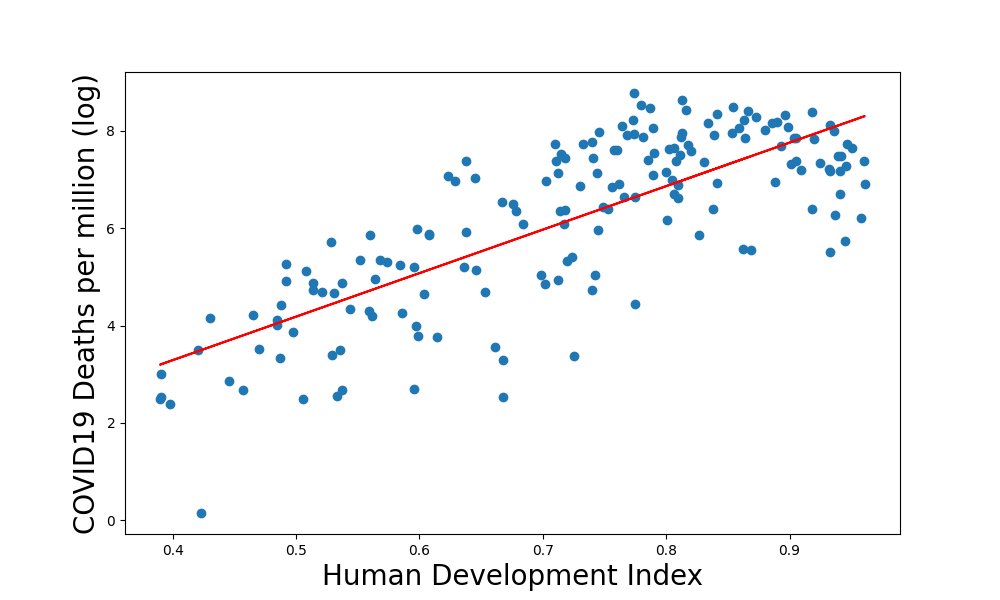}
        \caption{HDI}
        \label{fig6:third_image}
    \end{subfigure}

    \vspace{1em}

    % Third row: Two centered plots
    \makebox[\textwidth][c]{%
        \begin{subfigure}[t]{0.325\textwidth}
            \centering
            \includegraphics[width=\textwidth]{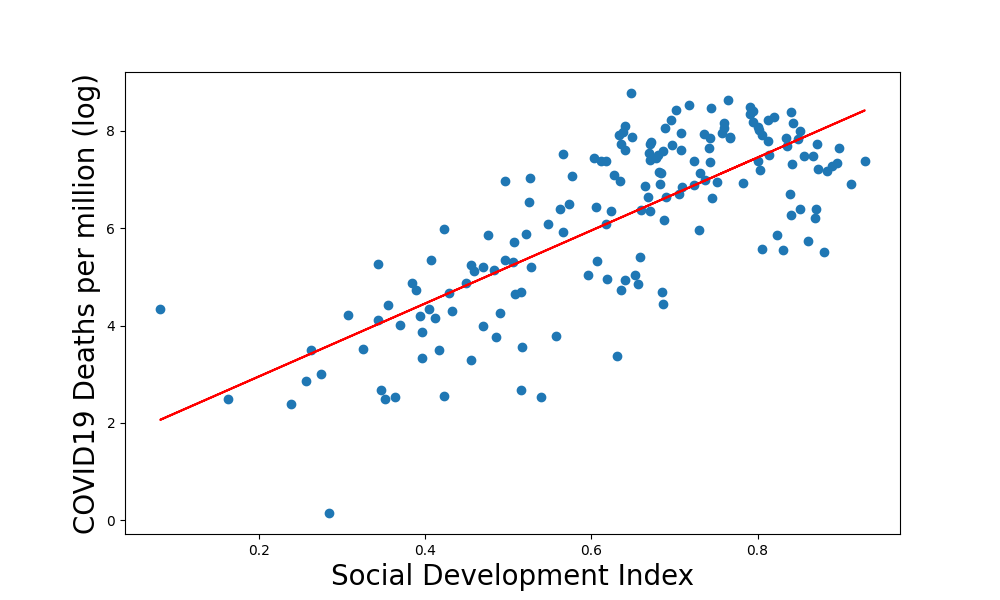}
            \caption{SDI}
            \label{fig6:plot_below4}
        \end{subfigure}
        \hspace{2em}
        \begin{subfigure}[t]{0.325\textwidth}
            \centering
            \includegraphics[width=\textwidth]{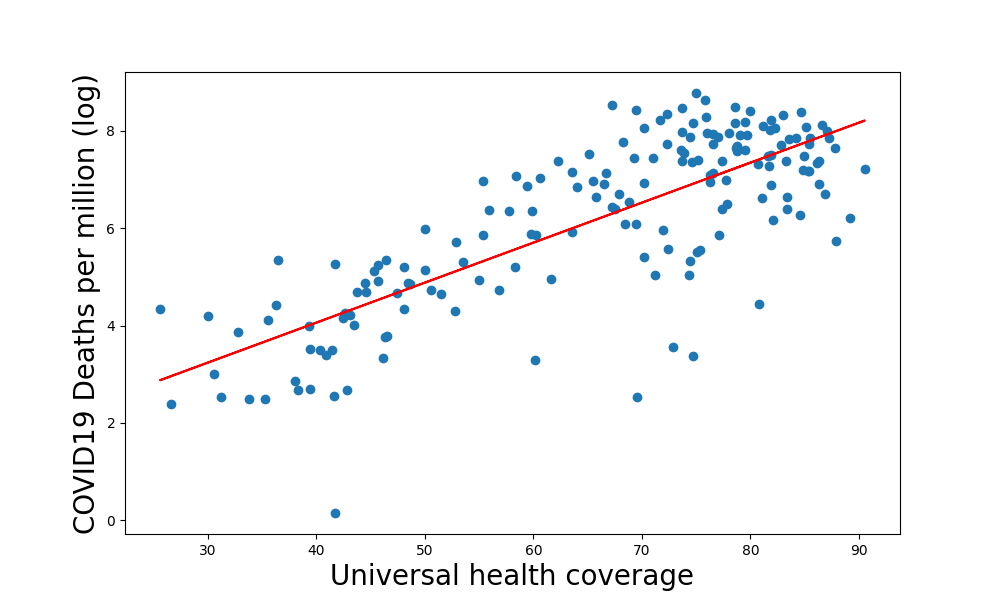}
            \caption{UHCI}
            \label{fig6:plot_below5}
        \end{subfigure}
    }

    \caption{Bivariate relationships between log–COVID-19 deaths per million and conventional indicators}
    \label{fig:deaths}
\end{figure}

To benchmark PoPStat–COVID19 against conventional socio-economic and demographic metrics, we computed Pearson correlations between each indicator and log-transformed COVID-19 cases and deaths per million.  Table~\ref{tab:popstat_covid19_full} summarises, for each indicator, the Pearson correlation coefficient, two-sided \(p\)-value, 95\% confidence interval, and variance explained (\(R^{2}\)).  The full scatterplots with ordinary-least-squares fits are shown in Figure~\ref{fig:cases} for cases and Figure~\ref{fig:deaths} for deaths, providing a visual comparison of slope steepness, dispersion, and outliers for each relationship.
\raggedbottom

\section{Discussion}
The COVID-19 pandemic revealed significant disparities in health impact and outcomes across countries, much of which can be attributed to differences in numerous socio-economic factors. The population pyramid of a country is a descriptive demographic statistic which visually represents a country’s age distribution. It represents the cumulative manifestation of various socioeconomic determinants, reflecting the healthcare system, social policy, fertility and mortality patterns, education levels, and demographic transition patterns. Therefore, it is reasonable to suggest that the population pyramid may be correlated with the impact of the COVID-19 pandemic. The existing literature confirms that the shape of the population pyramid strongly influences how each population faces the COVID-19 pandemic.~\cite{D_Manuel_2020}

This study proposes a mechanism to derive a statistical measure, PoPStat-COVID-19, which demonstrates a high correlation with COVID-19 severity by leveraging the demographic structure encoded in a country’s population pyramid.  This metric is dervied through Kullback-Leibler divergence-based comparison between a country's population distribution and an optimised reference pyramid. Thus, it quantifies the shape of the pyramid into a scalar variable, encapsulating how much a country's demographic profile diverges from an optimal structure. 

Our study demonstrates strong, statistically significant negative correlations between PoPDivergence and both COVID-19 incidence (r=-0.860, p$<$0.001) and mortality (r=-0.821, p$<$0.001), with Malta as the reference country. This indicates that countries whose population structures deviate significantly from that of Malta, which is typically an old-skewed country, experienced markedly lower COVID-19 incidence and mortality.

 These results align with previous research indicating that the countries with regressive population pyramids, characterised by a larger proportion of elderly individuals, exhibited a higher severity of COVID-19 outcomes.~\cite{Bonanad_2020} This is expected given that advanced age is the strongest independent predictor of COVID-19 morbidity and mortality.~\cite{Kofahi_2022} Older populations tend to inherent higher rates of comorbidities and require intensive medical care.~\cite{Mueller_2020} This amplifies the strain on healthcare systems and raises the overall fatality risk.

Conversely, countries with progressive pyramids, where a larger proportion of the population is younger, especially in Africa, show a different dynamic. Although these countries tend to have rapid transmission among younger generations, the clinical burden remains lower with milder outcomes and lower hospitalisation rates.~\cite{D_Manuel_2020} Therefore, despite the potential for rapid viral spread, this population structure allows for a reduced overall burden on health systems, acting more or less as a demographic buffer.    

In our robustness analysis, all ten regressive references returned large negative PoPStat–COVID19 values.  Demographically similar countries—Malta, Poland, the United States, and Japan—illustrate this stability: despite minor differences in their age–sex distributions, their PoPStat–COVID19\(_{\mathrm{cases}}\) values varied by only ~0.03 and PoPStat–COVID19\(_{\mathrm{deaths}}\) by ~0.03.  This tight clustering confirms that, when the reference pyramid is old and constrictive, PoPStat–COVID19 is highly robust to modest shifts in age structure.

By contrast, progressive references produced weaker—but directionally consistent—positive coefficients (median \(r=0.46\) for cases, \(r=0.43\) for deaths).  This asymmetry arises because KL divergence weights age groups in proportion to their prevalence in the reference: an older reference amplifies the signal from high‐risk cohorts, whereas a younger reference attenuates it.  Importantly, PoPStat–COVID19 remained statistically significant (\(p<0.001\)) for all regressive references and for eight of the ten progressive references, underscoring the robustness of the demographic association.

Traditional pandemic modelling indicators, such as median age, HDI, and GDP per capita, oversimplify demographics and have significant limitations. The median age overlooks critical distributional features, such as the proportion of vulnerable elderly populations, resulting in different risks within populations with the same median~\cite{T_Prem_2020}. Although HDI shows a strong correlation with COVID-19 case rates (r = 0.90, $R^2$ = 0.80), this association stems largely from broad socioeconomic factors rather than direct biological vulnerability. Furthermore, case rates themselves are highly susceptible to under-counting due to reporting biases, especially in under-resourced settings~\cite{B_Pei_2021}. In contrast, PoPDivergence directly encodes structural-demographic vulnerability by capturing differences in age distributions, thereby providing a biologically meaningful explanation for the variation in COVID-19 burden. Consequently, PoPStat–COVID19, based on PoPDivergence and utilising complete pre-pandemic age data, reliably predicts both case rates (explaining 74\% of the variance) and death rates (explaining 67\% of the variance). Beyond COVID-19, PoPStat can be applied to future pandemics in which age-dependent risk is present, such as influenza or RSV(Respiratory Syncytial Virus), providing a ready-to-use metric for demographic preparedness assessments. Its strength lies in this direct encoding of demographic risk, setting it apart from broader development-based measures, such as the HDI while performing comparably.

\section{Conclusion}
We have introduced \emph{PoPStat–COVID19}, a novel scalar metric that captures the influence of detailed age–sex structure on cross‐national COVID-19 incidence and mortality.  By combining Kullback–Leibler divergence of full population pyramids with Pearson correlation tuning, PoPStat–COVID19 distills complex demographic geometry into a single, interpretable value.  When calibrated to Malta’s pyramid, it explains 74\% of variance in COVID-19 cases and 67\% in COVID-19 deaths per million.

Robustness checks across twenty alternative high-performing reference pyramids demonstrate that \emph{PoPStat–COVID19} preserves its strong, statistically significant correlations whenever the substitute reference exhibits a similarly old-skewed age structure. Comparative benchmarking against eight widely used socio-economic and demographic indicators shows that the demographic-structure signal isolated by \emph{PoPStat–COVID19} matches the explanatory power of the best traditional predictors and surpasses most others.

PoPStat–COVID19 offers several key advantages as a demographic vulnerability metric. First, it provides demographic granularity by leveraging the full age distribution of population pyramids, capturing critical structural features like skewness and the concentration of high-risk age groups that are missed by simpler metrics like median age. Second, it ensures cross-national comparability, demonstrating robustness across more than 180 countries regardless of the specific reference population used, as long as the reference maintains a similar age profile. Third, it exhibits strong predictive strength, explaining a greater proportion of variance in COVID-19 case and death rates than widely used indicators such as GDP per capita, health coverage indices, or median age, thereby offering a biologically grounded and statistically powerful tool for pandemic risk assessment.

%
% ---- Bibliography ----
%
\bibliographystyle{splncs04}
\bibliography{chapters/newreferences}
\end{document}